\documentclass[twocolumn,showpacs,preprintnumbers,amsmath,amssymb]{revtex4}

\usepackage{graphicx}
\usepackage{dcolumn}
\usepackage{bm}

\begin{document}


\title{Small scale behavior of financial data}

\author{Andreas P. Nawroth}
\author{Joachim Peinke}%
\affiliation{%
Institut f\"ur Physik, Carl-von-Ossietzky Universit\"at Oldenburg, D-26111 Oldenburg, Germany
}%

\date{\today}

\begin{abstract}
A new approach is presented to describe the change in the statistics of the log return distribution of financial data as a function of the timescale. To this purpose a measure is introduced, which quantifies the distance of a considered distribution to a reference distribution. The existence of a small timescale regime is demonstrated, which exhibits different properties compared to the normal timescale regime for timescales larger than one minute. This  regime seems  to be universal for individual stocks. It is shown that the existence of this small timescale regime is not dependent on the special choice of the distance measure or the reference distribution. These findings have important implications for risk analysis, in particular for the probability of extreme events.
\end{abstract}

\maketitle

\section{\label{Introduction} Introduction}

The origin of mathematical finance dates back to Bachelier's famous thesis \textit{Th\'eorie De La Sp\'eculation} (see \cite{courtault2000}). As a central point in this work a normal distribution was assumed for financial returns. This assumption was for several reasons later changed by other authors to a normal distribution for the log return $r$ \cite{osborne1959}. The log return $r$ is defined in the following way:
\begin{eqnarray}
r(\tau):= log(P(t+\tau))-log(P(t))
\label{eq_def_logret}
\end{eqnarray}
where $P(t)$ denotes the price of the investment at time $t$. For alternative distributions to the log return distribution we refer to \cite{fama1965, mandelbrot1963, clark1973, mantegna1995, bouchaud2001, castaing1990a}. Other attempts \cite{lux1999} focused on the mechanism that  may produce such distributions. There remains a general problem to determine the correct family of distributions, based on an appropriate underlying stochastic process, incorporating the dependence of the shape of the distribution on the timescale. 

In the following we focus on the distribution (or the so called probability density function - pdf), which is in general dependent on the value of the log return itself as well as on the considered timescale. The question of the dependence of the shape of the distribution on the time scale was already posed in \cite{osborne1959}. Considering changes of the form of distributions requires to distinguish between changes due to the mean value, due to the standard deviation, see e.g. \cite{cont1997}, and due to the shape, see e.g. \cite{plerou1999, ghashghaie96}. A discussion of the importance of risk measures like VaR and their connection to the underlying distribution can be found in  \cite{hull1998, farias2004}. 

When considering individual stocks, for very large time scales the normalized distribution is quite similar to a Gaussian distribution. For small timescales a Non-Gaussian fat-tailed distribution is obtained. An interesting question now arises. Is this transition from a fat-tailed distribution towards a Gaussian a smooth and uniform process? A general non-parametric method, utilizing a Fokker-Planck equation in timescale, has been proposed, which provides a general description of how the shape of the distribution evolves with changing timescale \cite{renner01b}.  Although this approach is very general, it is based on assumptions that are partially no longer fulfilled for very small time scales (typically smaller than several minutes). Therefore here a specific non-parametric approach is presented, which provides insight into timescales covering seconds and minutes.

\section{\label{Data} Data}

In this study tick-by-tick data sets are used, in order to cover timescales as small as possible. The financial data sets were provided by the Karlsruher Kapitalmarkt Datenbank (KKMDB) \cite{luedecke1998}. The data sets contain all transactions on IBIS and XETRA in the corresponding period. The data sets used in this study span from the beginning of 1993 till the end of 2003 and contain $3-4\cdot10^6$ data points. Only stocks with a continuous history of trading in this period are considered. Results are presented for the three stocks with the largest number of trades in this period. These three stocks are Bayer, Volkswagen(VW) and Allianz.  In order to investigate changes of the shape of the distribution, we analyze in general normalized distributions and therefore look at the normalized return variable $R$
\begin{eqnarray}
R =  \frac{r - \overline{r}}{\sqrt{\overline{r^2} - \overline{r}^2}},
\label{eq_def_R}
\end{eqnarray}
where the average is taken over the whole data set. In order to compare the findings for stocks to other systems, the same analysis is performed for a turbulence data set. The data set was obtained by measuring the local longitudinal and transversal velocity component of a fluid in the turbulent wake behind a cylinder with a Taylor-based Reynolds number of 180 and contains $31 \cdot 10^6$ data points. For more details see \cite{siefert04}.

\section{Method}

A non-parametric approach to the detection of a change in shape of a distribution is a direct measurement of the distance between two distributions. $p_{ref}(R)$ denotes  the distribution for a reference  timescale and  $p(\tau,R)$ the distribution for another timescale. Firstly this allows verification of the frequently proposed assumption of a constant shape with respect to the timescale. Secondly if the shape is not constant this provides a quantitative measure of the size of the change in the shape of the distribution.  Therefore a measure is needed to quantify the distance between two distributions. Here, the Kullback-Leiber-Entropy  is used, which is defined as \cite{kullback1968}
\begin{eqnarray}
d_K(p(\tau,R),p_{ref}(R)) := \int \limits^{+\infty}_{-\infty}dR \; p(\tau,R) \cdot \ln \left ( \frac{p(\tau,R)}{p_{ref}(R)} \right ).
\label{eq_distance_KL}
\end{eqnarray}
In order to demonstrate the independence of our results on the particular choice of the measure we also use the weighted mean square error in logarithmic space
\begin{eqnarray}
\label{eq_def_distance_MSE}
\lefteqn{d_M(p(\tau,R),p_{ref}(R)) :=  } \hspace{-0.0cm} \\
\nonumber \\
& & \frac{\int \limits^{+\infty}_{-\infty}dR \; (p(\tau,R)+p_{ref}(R)) (\ln p(\tau,R)-\ln p_{ref}(R))^2}{\int \limits^{+\infty}_{-\infty}dR \; (p(\tau,R)+p_{ref}(R)) (\ln^2 p(\tau,R)+\ln^2 p_{ref}(R))} \nonumber .
\end{eqnarray}
Furthermore the chi-square distance is used as a third measure
\begin{eqnarray}
d_C(p(\tau,R),p_{ref}(R)) := \frac{ \int \limits^{+\infty}_{-\infty}dR \; (p(\tau,R)-p_{ref}(R))^2}{ \int \limits^{+\infty}_{-\infty}dR \; p_{ref}(R)}.
\label{eq_distance_Chi}
\end{eqnarray}
Using these distance measures it is possible to determine the distance of a log return distribution calculated for a certain timescale from a reference distribution.

\section{Evidence of a new universal small timescale regime}

\begin{figure}
\includegraphics[width= 5.5cm]{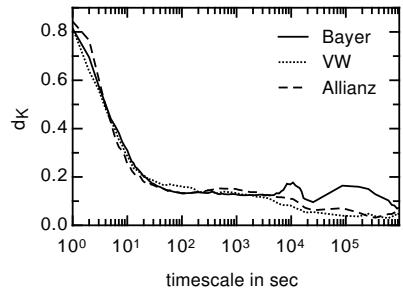}
\caption{\label{fig_distance_gauss_finance} Kullback-Leiber distance to the Gaussian distribution for three stocks.}
\end{figure}

For very large timescales the distribution is quite close to a Gaussian, therefore the Gaussian distribution is taken as reference distribution. In Fig. \ref{fig_distance_gauss_finance}
the Kullback-Leiber distance to the Gaussian distribution for three individual stocks is shown. It is evident, that the behavior changes considerably for timescales smaller than 100s. For such small scales the pdfs of financial data are considerably different from the Gaussian distribution. 

\begin{figure*}
\includegraphics[width= 16cm]{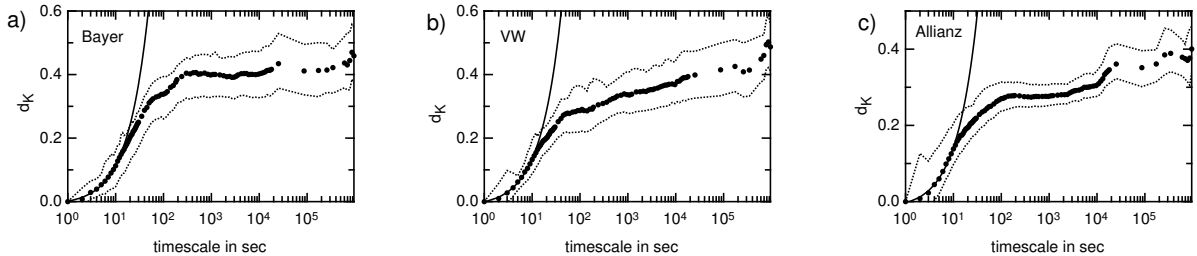}
\caption{\label{fig_dK_finance} The distance measure $d_K$ for a reference distribution $p_{ref}(R):=p(\tau=1s,R)$ for the individual stocks. The dots represent the estimated value, the dotted lines the one sigma error bound and the solid line  the linear fit for the first region.}
\end{figure*}

In a second step the distribution of the smallest scale of the considered asset is chosen as a reference distribution. In Fig. \ref{fig_dK_finance} 
 the distance $d_K$ to the smallest timescale for the three stocks is shown, together with the one sigma error (dotted lines). The  error estimate was  calculated by means of sub-samples of the data set to estimate the distribution of the distance measure. Again a transition behavior is seen, indicating a change in the stochastic behavior in the region 10s -100s. In all three cases the first region may be characterized by a linear increase of the distance measure $d_K$. The linear fit for this first region is drawn as a solid line in Fig. \ref{fig_dK_finance} (Note the use of semilog plots). 
 
\begin{figure*}
\includegraphics[width= 16cm]{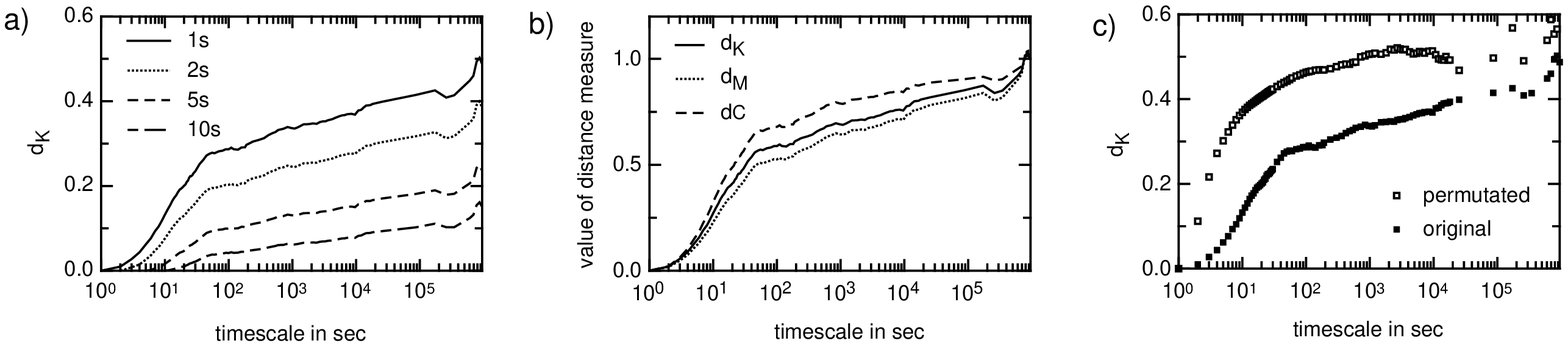}
\caption{\label{fig_dK_vw_verification}a) The distance measure $d_K$ for Volkswagen. b) Three different distance measures with $p_{ref}(R)=p(\tau=1s,R)$ for Volkswagen. c) Comparison of the distance measure $d_K$ with $p_{ref}(R)=p(\tau=1s,R)$ for the original and the permuted Volkswagen data set.}
\end{figure*}

In order to verify if the region displaying linear behavior is dependent on the chosen reference timescale, the analysis has been redone for different reference timescales. As an illustration, the results for Volkswagen are shown in Fig. \ref{fig_dK_vw_verification}a. For all these reference distributions the extent of the linear region (more precisely the upper bound) does not change. This and similar results for the other assets indicate, that the linear region is independent of the timescale that was chosen for the reference distribution. Next we discuss the influence of different distance measures (Eqs. (\ref{eq_distance_KL})-(\ref{eq_distance_Chi})). As an example the distance to the smallest timescale for VW is shown in Fig. \ref{fig_dK_vw_verification}b. Similar results were obtained for other stocks. For comparison all distance  measures were rescaled to the interval $[0,1]$ in Fig. \ref{fig_dK_vw_verification}b. For all three distance measures a division of the timescale in two parts characterized by the different functional behavior in these parts is evident. 

\begin{figure*}
\includegraphics[width= 13cm]{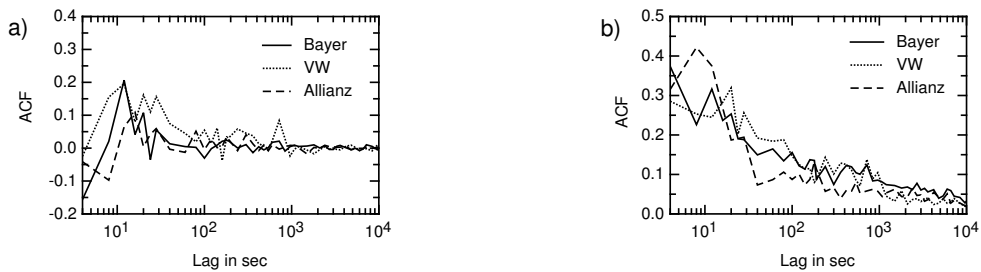}
\caption{\label{fig_acf_finance} a) Autocorrelation function of the log returns for three individual stocks. b) Autocorrelation function of the magnitude of the log returns for three individual stocks. }
\end{figure*}

A possible reason for the existence of different domains may be based on a specific relationship between consecutive increments on different timescales. One way to analyze this is to destroy all possible causal relationships of consecutive increments. This can be done by permuting all increments on a certain timescale (here the timescale of the reference distribution) and thereby creating a new time series with the same $p_{ref}(R)$. This new time series exhibits for small timescales a logarithmic increase in the distance measure $d_K$, see Fig. \ref{fig_dK_vw_verification}c. Further there is no longer a division into two distinct timescale intervals with different functional behavior of the distance measure $d_K$.
It is therefore evident, that the small timescale regime is due to functional relationships between consecutive increments. In order to investigate if the dependencies are linear, the autocorrelation function (ACF) of the non-uniformly sampled time series is calculated, cf. \cite{benedict2000}. The estimator for the autocorrelation is defined in the following way
\begin{eqnarray}
\lefteqn{\hat{\rho}(\tau, \Delta \tau) :=  \Bigg [ \sum_{i=1}^N \sum_{j=1}^N \big [r(\tau, t_i)-\bar{r}(\tau) \big ]} \hspace{1cm} \\
	& & \times \; \big [ r(\tau, t_j)-\bar{r}(\tau)\big ] b(t_j - t_i) \Bigg ] \nonumber \\
	& & \times \Bigg [ \Bigg [ \sum_{i=1}^N \sum_{j=1}^N \big [ r^2(\tau, t_i) - \bar{r}^2(\tau) \big ] b(t_j - t_i) \Bigg ] \nonumber \\
	& & \times \Bigg [ \sum_{i=1}^N \sum_{j=1}^N \big [ r^2(\tau, t_j) - \bar{r}^2(\tau) \big ] b(t_j - t_i) \Bigg ] \Bigg ]^{-\frac{1}{2}} \nonumber
\label{eq_autocorrelation}
\end{eqnarray}
\begin{eqnarray}
b(t_j - t_i) = \left \{  \begin{array} {r@{\quad }l}
				1 & for \quad |(t_j -t_i) - \Delta \tau| < \delta \Delta \tau \\
				0 & otherwise
			\end{array} \right.
\label{eq_autocorrelation_mean}
\end{eqnarray}
where $r(\tau, t_i)$ is the log return on the timescale $\tau$ at the time $t_i$ and $\delta$ a small number. The results for the ACF, computed on a timescale of four seconds, are shown in Fig. \ref{fig_acf_finance}a+b. The computation of the ACF for smaller timescales becomes increasingly difficult due to the very small number of available log returns. In agreement with the literature \cite{bouchaud2001} \cite{dacorogna2001}, there is a negative autocorrelation for the smallest lag, while for larger lags the ACF yields values very close to zero. The ACF of the magnitude of the log returns is considered in Fig. \ref{fig_acf_finance}b. Here there is a strong positive autocorrelation for the smallest lag, which slowly  decays for larger lags. However, for both ACFs and all the considered stocks there is no indication of a small timescale regime in the ACF.  It therefore appears that the functional relationship between consecutive increments, which causes the small timescale regime, is of nonlinear nature.

\section{Comparison with turbulence data}

\begin{figure*}
\includegraphics[width= 15cm]{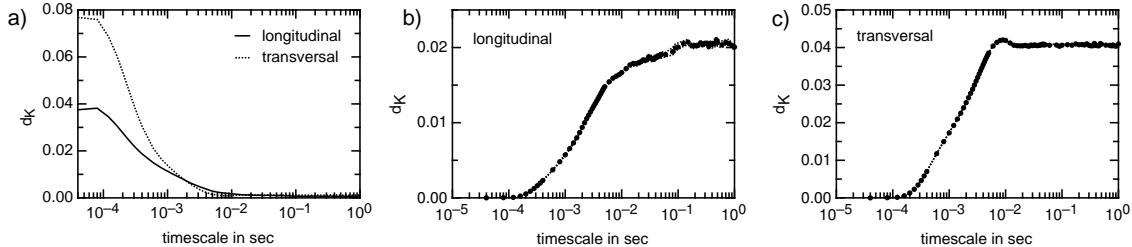}
\caption{\label{fig_dK_WK2808} Turbulence data: a) The dependence of $d_K$ to the Gaussian distribution on the timescale. b+c) The distance measure $d_K$ for a reference distribution $p_{ref}(R):=p(\tau=4 \cdot 10^{-5}s,R)$. The dots represent the estimated value, the dotted lines the one sigma error bound.}
\end{figure*}

In \cite{ghashghaie96} and \cite{mantegna1997} it has been shown, that finance and  turbulence data display common properties. The analysis described above is therefore also performed  with turbulence data in order to see if a small timescale regime is present in that case as well. In Fig. \ref{fig_dK_WK2808}
the distances $d_K$ of the distribution of the velocity increments with respect to the Gaussian distribution (a) and with respect to a small scale reference distribution (b and c) for the turbulence data are shown. The qualitative behavior for larger timescales is similar to that observed for individual stocks, while for  smaller timescales  the behavior differs. It is important to note the difference in scale of the distance measure in Figs. \ref{fig_distance_gauss_finance} and \ref{fig_dK_WK2808}a.

\section{Applications}

\begin{figure*}
\includegraphics[width= 16cm]{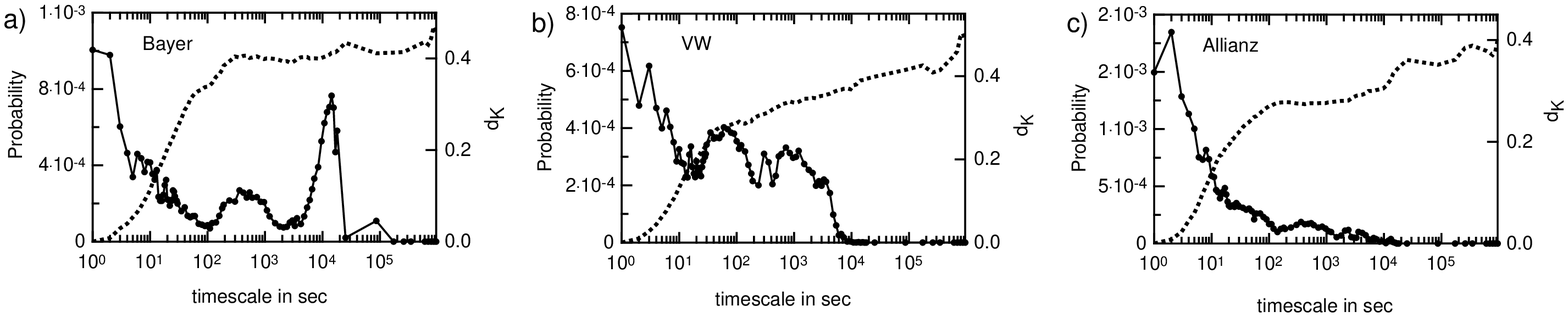}
\caption{\label{fig_cum_all} Comparison between the probability mass beyond the 10th standard deviation (solid line) and $d_K$ (dotted line).}
\end{figure*}

How does the specific behavior of the small timescale regime translate into practical applications? The deviation from the Gaussian distribution is increasing much faster in the small timescale regime than in the normal timescale regime. A visual inspection shows, that the considered log return distributions deviate into the direction of fat-tailed distributions. Therefore the probability mass in the tails of the distribution should increase faster by entering the small timescale regime. In order to analyze this, the probability mass in the tails of the distribution, i.e. the probability mass beyond the 10th standard deviation, where left and right tail are considered together, is calculated and the results are compared to the distance measure $d_K$. The reference timescale is one second. The results are shown in Fig. \ref{fig_cum_all}. 
In all three cases it is evident, that the change of the distance measure corresponds to a change of probability mass in the tails of the distribution. In the small timescale regime the increase in the probability mass in the tails of the distribution is very pronounced. The estimates of probability mass, for timescales larger than $10^3s$, are rather noisy, due to the effect that events are quite rare in this region. 

\section{Conclusions}

Summarizing, it has been demonstrated that the properties of the log return distribution of stocks do not change uniformly if one goes to smaller timescales. Instead, for small timescales a distinct regime is entered with different properties. In this small timescale regime, the shape of the distribution changes much faster than one would expect by extrapolating the behavior of the normal timescale regime. This small timescale regime extends for individual stocks to our knowledge from timescales of around 1s to timescales of around 15s. The small timescale regime can be characterized by a linear dependence of the Kullback-Leiber distance $d_K$ on the timescale, if as a reference distribution a log return distribution on a very small timescale is chosen. In the normal timescale regime the dependence is much slower and can be assumed to be logarithmic or for very large timescales independent of the timescale. This result seems to be independent of the chosen reference distribution as long as it is a log return distribution on a sufficient small timescale. If the Gaussian distribution is taken as a reference distribution, $d_K$ is rising very fast with decreasing timescale in the small timescale regime, while it stays nearly constant in the normal timescale regime in accordance with \cite{plerou1999}. This indicates a very fast deviation from a Gaussian-like shape in the small timescale regime. These results could be confirmed with different distance measures. Further it has been shown that this small timescale regime is a specific feature of the financial data investigated here. For turbulence data no such small timescale regime is observed, although financial and turbulence data sets exhibit similarities in the normal timescale regime. For very small timescales the shape of the distribution, in contrast to the findings for the financial data sets,  changes slower than one would expect by extrapolating the behavior of the normal timescale regime. Two prominent candidates for this effect are the dissipation on small scales for turbulence \cite{siefert04} and the noise added by the measurement system. Furthermore the particular small timescale regime for individual stocks cannot be reproduced by trivial randomized data. As an application of this new approach it has been demonstrated that on entering the small timescale regime, a large increase in the probability mass in the tails of the distribution occurs, which could lead to very different risk characteristics in comparison to that of larger timescales.





\newpage 
\bibliography{AG_LITERATUR}

\begin{thebibliography}{21}
\expandafter\ifx\csname natexlab\endcsname\relax\def\natexlab#1{#1}\fi
\expandafter\ifx\csname bibnamefont\endcsname\relax
  \def\bibnamefont#1{#1}\fi
\expandafter\ifx\csname bibfnamefont\endcsname\relax
  \def\bibfnamefont#1{#1}\fi
\expandafter\ifx\csname citenamefont\endcsname\relax
  \def\citenamefont#1{#1}\fi
\expandafter\ifx\csname url\endcsname\relax
  \def\url#1{\texttt{#1}}\fi
\expandafter\ifx\csname urlprefix\endcsname\relax\def\urlprefix{URL }\fi
\providecommand{\bibinfo}[2]{#2}
\providecommand{\eprint}[2][]{\url{#2}}

\bibitem[{\citenamefont{Courtault et~al.}(2000)\citenamefont{Courtault,
  Kabanov, Bru, Crep\'el, Lebon, and Marchand}}]{courtault2000}
\bibinfo{author}{\bibfnamefont{J.}~\bibnamefont{Courtault}},
  \bibinfo{author}{\bibfnamefont{Y.}~\bibnamefont{Kabanov}},
  \bibinfo{author}{\bibfnamefont{B.}~\bibnamefont{Bru}},
  \bibinfo{author}{\bibfnamefont{P.}~\bibnamefont{Crep\'el}},
  \bibinfo{author}{\bibfnamefont{I.}~\bibnamefont{Lebon}}, \bibnamefont{and}
  \bibinfo{author}{\bibfnamefont{A.~L.} \bibnamefont{Marchand}},
  \bibinfo{journal}{Mathematical Finance} \textbf{\bibinfo{volume}{10}},
  \bibinfo{pages}{341} (\bibinfo{year}{2000}).

\bibitem[{\citenamefont{Osborne}(1959)}]{osborne1959}
\bibinfo{author}{\bibfnamefont{M.}~\bibnamefont{Osborne}},
  \bibinfo{journal}{Operations Research} \textbf{\bibinfo{volume}{7}},
  \bibinfo{pages}{145} (\bibinfo{year}{1959}).

\bibitem[{\citenamefont{Fama}(1965)}]{fama1965}
\bibinfo{author}{\bibfnamefont{E.}~\bibnamefont{Fama}},
  \bibinfo{journal}{Journal of Business} \textbf{\bibinfo{volume}{38}},
  \bibinfo{pages}{34} (\bibinfo{year}{1965}).

\bibitem[{\citenamefont{Mandelbrot}(1963)}]{mandelbrot1963}
\bibinfo{author}{\bibfnamefont{B.}~\bibnamefont{Mandelbrot}},
  \bibinfo{journal}{The Journal of Business} \textbf{\bibinfo{volume}{36}},
  \bibinfo{pages}{394} (\bibinfo{year}{1963}).

\bibitem[{\citenamefont{Clark}(1973)}]{clark1973}
\bibinfo{author}{\bibfnamefont{P.~K.} \bibnamefont{Clark}},
  \bibinfo{journal}{Econometrica} \textbf{\bibinfo{volume}{41}},
  \bibinfo{pages}{135} (\bibinfo{year}{1973}).

\bibitem[{\citenamefont{Mantegna and Stanley}(1995)}]{mantegna1995}
\bibinfo{author}{\bibfnamefont{R.~N.} \bibnamefont{Mantegna}} \bibnamefont{and}
  \bibinfo{author}{\bibfnamefont{H.~E.} \bibnamefont{Stanley}},
  \bibinfo{journal}{Nature} \textbf{\bibinfo{volume}{376}}, \bibinfo{pages}{46}
  (\bibinfo{year}{1995}).

\bibitem[{\citenamefont{Bouchaud and Potters}(2001)}]{bouchaud2001}
\bibinfo{author}{\bibfnamefont{J.~P.} \bibnamefont{Bouchaud}} \bibnamefont{and}
  \bibinfo{author}{\bibfnamefont{M.}~\bibnamefont{Potters}},
  \emph{\bibinfo{title}{Theory of Financial Risks}}
  (\bibinfo{publisher}{Cambridge University Press}, \bibinfo{year}{2001}).

\bibitem[{\citenamefont{Castaing et~al.}(1990)\citenamefont{Castaing, Gagne,
  and Hopfinger}}]{castaing1990a}
\bibinfo{author}{\bibfnamefont{B.}~\bibnamefont{Castaing}},
  \bibinfo{author}{\bibfnamefont{Y.}~\bibnamefont{Gagne}}, \bibnamefont{and}
  \bibinfo{author}{\bibfnamefont{E.~J.} \bibnamefont{Hopfinger}},
  \bibinfo{journal}{Physica D} \textbf{\bibinfo{volume}{46}},
  \bibinfo{pages}{177} (\bibinfo{year}{1990}).

\bibitem[{\citenamefont{Lux and Marchesi}(1999)}]{lux1999}
\bibinfo{author}{\bibfnamefont{T.}~\bibnamefont{Lux}} \bibnamefont{and}
  \bibinfo{author}{\bibfnamefont{M.}~\bibnamefont{Marchesi}},
  \bibinfo{journal}{Nature} \textbf{\bibinfo{volume}{397}},
  \bibinfo{pages}{498} (\bibinfo{year}{1999}).

\bibitem[{\citenamefont{Cont et~al.}(1997)\citenamefont{Cont, Potters, and
  Bouchaud}}]{cont1997}
\bibinfo{author}{\bibfnamefont{R.}~\bibnamefont{Cont}},
  \bibinfo{author}{\bibfnamefont{M.}~\bibnamefont{Potters}}, \bibnamefont{and}
  \bibinfo{author}{\bibfnamefont{J.~P.} \bibnamefont{Bouchaud}},
  \bibinfo{journal}{Proc. CNRS Workshop on Scale Invariance}
  (\bibinfo{year}{1997}), \urlprefix\url{cond-mat/9607120}.

\bibitem[{\citenamefont{Plerou et~al.}(1999)\citenamefont{Plerou, Gopikrishnan,
  Amaral, Meyer, and Stanley}}]{plerou1999}
\bibinfo{author}{\bibfnamefont{V.}~\bibnamefont{Plerou}},
  \bibinfo{author}{\bibfnamefont{P.}~\bibnamefont{Gopikrishnan}},
  \bibinfo{author}{\bibfnamefont{L.~A.~N.} \bibnamefont{Amaral}},
  \bibinfo{author}{\bibfnamefont{M.}~\bibnamefont{Meyer}}, \bibnamefont{and}
  \bibinfo{author}{\bibfnamefont{H.~E.} \bibnamefont{Stanley}},
  \bibinfo{journal}{Phys. Rev. E} \textbf{\bibinfo{volume}{60}},
  \bibinfo{pages}{6519} (\bibinfo{year}{1999}).

\bibitem[{\citenamefont{Ghashghaie et~al.}(1996)\citenamefont{Ghashghaie,
  Breymann, Peinke, Talkner, and Dodge}}]{ghashghaie96}
\bibinfo{author}{\bibfnamefont{S.}~\bibnamefont{Ghashghaie}},
  \bibinfo{author}{\bibfnamefont{W.}~\bibnamefont{Breymann}},
  \bibinfo{author}{\bibfnamefont{J.}~\bibnamefont{Peinke}},
  \bibinfo{author}{\bibfnamefont{P.}~\bibnamefont{Talkner}}, \bibnamefont{and}
  \bibinfo{author}{\bibfnamefont{Y.}~\bibnamefont{Dodge}},
  \bibinfo{journal}{Nature} \textbf{\bibinfo{volume}{381}},
  \bibinfo{pages}{767} (\bibinfo{year}{1996}).

\bibitem[{\citenamefont{Hull and White}(1998)}]{hull1998}
\bibinfo{author}{\bibfnamefont{J.}~\bibnamefont{Hull}} \bibnamefont{and}
  \bibinfo{author}{\bibfnamefont{A.}~\bibnamefont{White}},
  \bibinfo{journal}{Journal of Derivatives} \textbf{\bibinfo{volume}{5}},
  \bibinfo{pages}{9} (\bibinfo{year}{1998}).

\bibitem[{\citenamefont{Farias et~al.}(2004)\citenamefont{Farias, Ornelas, and
  Barbachan}}]{farias2004}
\bibinfo{author}{\bibfnamefont{A.}~\bibnamefont{Farias}},
  \bibinfo{author}{\bibfnamefont{J.}~\bibnamefont{Ornelas}}, \bibnamefont{and}
  \bibinfo{author}{\bibfnamefont{J.}~\bibnamefont{Barbachan}},
  \bibinfo{journal}{Stochastic Finance 2004}  (\bibinfo{year}{2004}).

\bibitem[{\citenamefont{Renner et~al.}(2001)\citenamefont{Renner, Peinke, and
  Friedrich}}]{renner01b}
\bibinfo{author}{\bibfnamefont{C.}~\bibnamefont{Renner}},
  \bibinfo{author}{\bibfnamefont{J.}~\bibnamefont{Peinke}}, \bibnamefont{and}
  \bibinfo{author}{\bibfnamefont{R.}~\bibnamefont{Friedrich}},
  \bibinfo{journal}{Physica A} \textbf{\bibinfo{volume}{298}},
  \bibinfo{pages}{499} (\bibinfo{year}{2001}).

\bibitem[{\citenamefont{L\"udecke}(1998)}]{luedecke1998}
\bibinfo{author}{\bibfnamefont{T.}~\bibnamefont{L\"udecke}},
  \bibinfo{journal}{Discussion Paper No. 190 University of Karlsruhe}
  (\bibinfo{year}{1998}).

\bibitem[{\citenamefont{Siefert and Peinke}(2004)}]{siefert04}
\bibinfo{author}{\bibfnamefont{M.}~\bibnamefont{Siefert}} \bibnamefont{and}
  \bibinfo{author}{\bibfnamefont{J.}~\bibnamefont{Peinke}},
  \bibinfo{journal}{Phys. Rev. E} \textbf{\bibinfo{volume}{70}},
  \bibinfo{pages}{015302} (\bibinfo{year}{2004}).

\bibitem[{\citenamefont{Kullback}(1968)}]{kullback1968}
\bibinfo{author}{\bibfnamefont{S.}~\bibnamefont{Kullback}},
  \emph{\bibinfo{title}{Information Theory And Statistics}}
  (\bibinfo{publisher}{Dover Publications}, \bibinfo{year}{1968}).

\bibitem[{\citenamefont{Benedict et~al.}(2000)\citenamefont{Benedict, Nobach,
  and Tropea}}]{benedict2000}
\bibinfo{author}{\bibfnamefont{L.~H.} \bibnamefont{Benedict}},
  \bibinfo{author}{\bibfnamefont{H.}~\bibnamefont{Nobach}}, \bibnamefont{and}
  \bibinfo{author}{\bibfnamefont{C.}~\bibnamefont{Tropea}},
  \bibinfo{journal}{Meas. Sci. Technol.} \textbf{\bibinfo{volume}{11}},
  \bibinfo{pages}{1089} (\bibinfo{year}{2000}).

\bibitem[{\citenamefont{Dacorogna et~al.}(2001)\citenamefont{Dacorogna,
  Gen\c{c}ay, M\"uller, Olsen, and Pictet}}]{dacorogna2001}
\bibinfo{author}{\bibfnamefont{M.~M.} \bibnamefont{Dacorogna}},
  \bibinfo{author}{\bibfnamefont{R.}~\bibnamefont{Gen\c{c}ay}},
  \bibinfo{author}{\bibfnamefont{U.}~\bibnamefont{M\"uller}},
  \bibinfo{author}{\bibfnamefont{R.~B.} \bibnamefont{Olsen}}, \bibnamefont{and}
  \bibinfo{author}{\bibfnamefont{O.~V.} \bibnamefont{Pictet}},
  \emph{\bibinfo{title}{An Introduction to High-Frequency Finance}}
  (\bibinfo{publisher}{Academic Press}, \bibinfo{year}{2001}).

\bibitem[{\citenamefont{Mantegna and Stanley}(1997)}]{mantegna1997}
\bibinfo{author}{\bibfnamefont{R.}~\bibnamefont{Mantegna}} \bibnamefont{and}
  \bibinfo{author}{\bibfnamefont{H.}~\bibnamefont{Stanley}},
  \bibinfo{journal}{Physica A} \textbf{\bibinfo{volume}{239}},
  \bibinfo{pages}{255} (\bibinfo{year}{1997}).

\end{thebibliography}

\end{document}